\documentclass[superscriptaddress, aps, showpacs, floatfix, prl, twocolumn]{revtex4-2}

\usepackage{graphicx}% Include figure files
\usepackage{dcolumn}% Align table columns on decimal point
\usepackage{bm}% bold math
\usepackage{xcolor}

\def\LCPO/{LiCoPO$_4$}

\begin{document}

\title{Imaging antiferromagnetic domains in \LCPO/ via the optical magnetoelectric effect}

\author{B.~T{\'o}th}
\email{toth.boglarka@ttk.bme.hu}
\affiliation{Department of Physics, Institute of Physics, Budapest University of Technology and Economics, M\H{u}egyetem rkp. 3., H-1111 Budapest, Hungary}

\author{V.~Kocsis}
\affiliation{RIKEN Center for Emergent Matter Science (CEMS), Wako, Saitama 351-0198, Japan}
\affiliation{Institut f\"ur Festk\"orperforschung, Leibniz IFW Dresden, 01069 Dresden, Germany}

\author{Y.~Tokunaga}
\affiliation{RIKEN Center for Emergent Matter Science (CEMS), Wako, Saitama 351-0198, Japan}
\affiliation{Department of Advanced Materials Science, University of Tokyo, Kashiwa 277-8561, Japan}

\author{Y.~Taguchi}
\affiliation{RIKEN Center for Emergent Matter Science (CEMS), Wako, Saitama 351-0198, Japan}

\author{Y.~Tokura}
\affiliation{RIKEN Center for Emergent Matter Science (CEMS), Wako, Saitama 351-0198, Japan}
%\affiliation{Quantum-Phase Electronics Center, Department of Applied Physics, University of Tokyo, Tokyo 113-8656, Japan}
\affiliation{Tokyo College, University of Tokyo, Hongo, Tokyo 113-8656, Japan}
\affiliation{Department of Applied Physics, University of Tokyo, Hongo, Tokyo 113-8656, Japan}

\author{S.~Bord{\'a}cs}
\affiliation{Department of Physics, Institute of Physics, Budapest University of Technology and Economics, M\H{u}egyetem rkp. 3., H-1111 Budapest, Hungary}
\affiliation{HUN-REN–BME Condensed Matter Physics Research Group, Budapest University of Technology and Economics, M\H{u}egyetem rkp. 3., H-1111 Budapest, Hungary}
\affiliation{Experimental Physics V, Center for Electronic Correlations and Magnetism, University of Augsburg, D-86135 Augsburg, Germany}

\begin{abstract}

Antiferromagnetic (AFM) materials are considered as promising building blocks of novel data storage devices, still, detecting and manipulating AFM domains have remained challenging. Here, we demonstrate that the two antiphase domains of the magnetoelectric antiferromagnet \LCPO/ can be distinguished by their light absorption difference. Using visible and infrared spectroscopy, we observed spontaneous non-reciprocal absorption, also termed as directional dichroism, at the crystal field excitations of Co$^{2+}$ ions coordinated by distorted oxygen octahedra. This absorption contrast is particularly pronounced near the telecommunication wavelength of 1550\,nm. These findings allowed us to image the AFM domains in \LCPO/ using a simple transmission light microscopy setup. Our findings suggest that optical magnetoelectric effects offer promising routes for probing the AFM order parameter in non-centrosymmetric transition metal compounds.

\end{abstract}

\maketitle

The detection of antiferromagnetic (AFM) domains with no net magnetization has been a challenge ever since the discovery of antiferromagnets \cite{Neel1972, Shull1994}. The recent progress of AFM spintronics \cite{Wadley2016} gave a new boost to the study of AFM materials. The application of such compounds could have significant benefits as 1) the rich variety of AFM orders offers new possibilities to encode information, 2) their THz dynamics promise great speed and 3) the absence of net magnetization makes them robust against stray magnetic fields \cite{Jungwirth2016, Baltz2018}. Despite all efforts, however, the electric or optical detection of the AFM domains is still difficult \cite{Jungwirth2016, Baltz2018, Nemec2018}.

When the magnetic order simultaneously breaks both the time-reversal and the inversion symmetries the magnetoelectric (ME) effect becomes allowed, namely electric fields, $\mathbf{E}$ can induce magnetization, $\mathbf{M}$ and magnetic fields, $\mathbf{H}$ can generate electric polarization, $\mathbf{P}$. The measurement of the ME susceptibility, $\chi_{ij}$ provides a rare opportunity to detect not only the orientation (director) but also the sign of the AFM order parameter, $\mathbf{L}$ -- the difference of the sublattice magnetizations -- as demonstrated e.g.~in the prototypical ME compound Cr$_2$O$_3$ \cite{Landau1960book, Astrov1960, Astrov1961, Folen1961, Rado1961}. 
Moreover, intriguing optical effects can also emerge at the mixed electric and magnetic dipole excitations of these compounds, which opens further possibilities to detect the AFM order and image its domains. Indeed, AFM domains were recently observed by imaging the spatial variation of the nonreciprocal linear dichroism in Pb(TiO)Cu$_4$(PO$_4$)$_4$ \cite{Kimura2020} and the electric field induced magnetic circular dichroism (MCD) in Cr$_2$O$_3$ \cite{Hayashida2022}.

\begin{figure}[h]
\centering
\includegraphics[width = \linewidth]{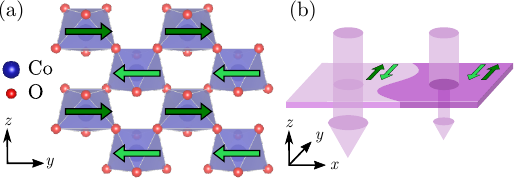}
%\captionsetup{justification=Justified}
\caption{ (a) Crystal and magnetic structure of \LCPO/ viewed from the $x$ direction. Green arrows show the direction of magnetic moments, their arrangement corresponds to one of the antiferromagnetic domains.  (b) Schematic illustration of the absorption contrast between the antiphase antiferromagnetic (AFM) domains observed on a thin $xy$ cut of \LCPO/.}
\label{fig:fig_concept}
\end{figure}

Besides the above-mentioned light polarization sensitive optical anisotropies, ME compounds also show the so-called nonreciprocal directional dichroism (NDD) \cite{Jung2004, Saito2008, Saito2008PRL, Bordacs2012, Kezsmarki2015, Kocsis2018, Sato2020, Vit2021, Kimura2024}, which is the absorption difference for counterpropagating beams ($\pm$$\mathbf{k}$), present even for unpolarized light \cite{Barron2004book, Szaller2013}. As the strongly and weakly absorbing directions are reversed for the domains having time-reversed magnetic states, NDD can be utilized for imaging such ME AFM domains as shown in Fig.~\ref{fig:fig_concept}(b) \cite{Saito2009, Sato2022}. The advantage of this method is its simplicity. Compared to other imaging techniques \cite{Nemec2018}, it neither requires strong lasers as second harmonics generation (SHG) microscopy nor polarization-sensitive elements. A prerequisite of NDD-based light microscopy is the presence of finite NDD in the visible (VIS) or near-infrared (NIR) spectral ranges which has been reported in CuB$_2$O$_4$ \cite{Saito2008, Saito2008PRL}, MnTiO$_3$ \cite{Sato2020} and LiNiPO$_4$ \cite{Kimura2024}.

In this Letter, we demonstrate by broadband infrared and VIS absorption spectroscopy that the ME antiferromagnet \LCPO/ shows large NDD at the crystal field excitation of Co$^{2+}$ ions. By selectively stabilizing one of the two time-reversed (or antiphase) domains, we detected significantly different absorption for the two domains. Particularly, the absorption contrast is as high as 34\,\% at around the telecommunication wavelength 1550\,nm (0.8\,eV). With a simple spatial laser scanning setup, we mapped the light intensity transmitted through a zero-field-cooled (ZFC) sample and succeeded in visualizing the AFM domains of \LCPO/ by exploiting their spontaneous NDD.

At room temperature, \LCPO/ crystallizes in the orthorhombic olivine structure (space group: $Pnma$). The magnetic Co$^{2+}$ ions coordinated by distorted oxygen octahedra occupy sites having only a mirror plane symmetry (Wyckoff position: 4h). The low site symmetry allows a local electric dipole lying within the mirror plane, but their sum over the unit cell vanishes \cite{Kocsis2018}. Below $T_{\mathrm{N}}$\,=\,21.8\,K, the S\,=\,3/2 spins of the Co$^{2+}$ ions order into a two-sublattice collinear AFM structure with moments parallel to the $y$ axis (see Fig.~\ref{fig:fig_concept}a) \cite{Santoro1966}. This structure is consistent with the magnetic space group $Pnma'$. In this symmetry, the time-reversal and the inversion symmetries are simultaneously broken, thus, the ME effect is allowed \cite{Rivera1994, Ederer2007}. % In fact, values of the two non-vanishing components of the linear ME susceptibility are among the largest reported so far: $\chi_{xy}/c$\,=\,15\,ps/m and $\chi_{yx}/c$\,=\,32\,ps/m, where $c$ is the speed of light \cite{Rivera1994, Fiebig2005}. 
The values of the two non-vanishing components of the linear ME susceptibility are rather large: $\chi_{xy}/c$\,=\,15\,ps/m and $\chi_{yx}/c$\,=\,32\,ps/m, where $c$ is the speed of light \cite{Rivera1994, Fiebig2005}. The emergence of the ME effect is explained by others with the non-zero ferrotoroidal moment allowed in $Pnma'$ \cite{Ederer2007, Zimmermann2014,Kocsis2019}, which may appear as the sum of the cross-products of the local electric and magnetic moments remaining non-vanishing despite the lack of net polarization and magnetization \cite{Kocsis2018, Kocsis2019}. Although neutron scattering experiments revealed a small, 4.6$^\circ$ rotation of the spins away from the $y$ axis \cite{Vaknin2002}, a later calculation found that the additional components of the ME tensor induced by the lower symmetry are small, thus, neglected in the literature \cite{Ederer2007, Zimmermann2014}. Via the linear ME susceptibility, the free energy of one of the AFM domains can be lowered compared to the other, when both of the corresponding electric and magnetic fields are applied. Therefore, a single AFM domain can be stabilized by cooling a \LCPO/ sample in these fields across the N\'eel temperature, $T_{\mathrm{N}}$. This procedure is referred to as ME poling in the literature. Recent experiments even demonstrated isothermal switching of the AFM domains close to $T_{\mathrm{N}}$ \cite{Kocsis2021}.

\begin{figure}[h]
\centering
\includegraphics[width = \linewidth]{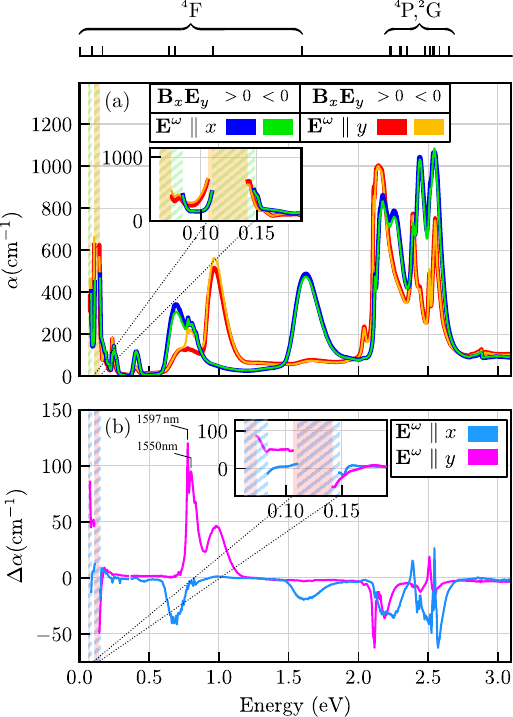}
%\captionsetup{justification=Justified}
\caption{ (a) Absorption spectra of \LCPO/ measured at 5\,K, after magnetoelectric (ME) poling for light polarizations $E^{\omega}\parallel x$ (blue, green) and $E^{\omega}\parallel y$ (red, orange). An AFM domain is selected depending on the sign of ME free energy term $B_xE_y$, where $B_x$ and $E_y$ are the poling magnetic and electric fields. In the colored energy ranges the sample is opaque, i.e.,~the absorption is beyond the detection limit. On the top of the panel, the energy levels are reproduced from the crystal field calculations of Ref.~\cite{Kornev1999}. (b) Absorption difference of the two AFM domains for different polarizations. The insets show a magnified energy range between 0.06 and 0.18\,eV.  }
\label{fig:fig_spectra}
\end{figure}

%%%%%%%%%%%
% Methods
%%%%%%%%%%%

We studied \LCPO/ single crystals, which were grown by the floating-zone method as described elsewhere \cite{Saint-Martin2008}. We measured the light intensity transmitted through a reference hole, $I_\mathrm{ref}$ and the sample, $I_\mathrm{samp}$ covering an identical hole with 700\,$\mu$m in diameter. Low temperatures were reached in an Oxford continuous flow optical cryostat, with a small permanent magnet fixed inside, for poling. We used a combination of a grating spectrometer (Newport Cornerstone 260) and a Fourier transform infrared spectrometer (Varian 670 FTIR) to cover a broad spectral range between 0.075\,eV and 3.1\,eV. We calculated the absorption spectra as  
\begin{equation}
    \alpha = -\frac{1}{d}\mathrm{ln}\left(\frac{I_{\mathrm{samp}}}{I_{\mathrm{ref}}}   \right),
\end{equation}
where $d$\,$\sim$\,60\,$\mu$m is the sample thickness. Reflection losses can be neglected as $n$ = 1.714 \cite{Rivera1994}. According to crystal symmetry, NDD appears for light propagation $\boldsymbol{k}\parallel z$ \cite{Kocsis2018}, thus, we studied an $xy$ cut with light beam propagating normal to the plane. As \LCPO/ is orthorhombic, we set the incident light polarization parallel to either $\mathbf{E}^{\omega}\parallel x$ or $\mathbf{E}^{\omega}\parallel y$. We applied poling fields with magnitude $\mu_0\mathbf{H}=$$\pm$$200$\,mT and $\mathbf{E}=$$\pm$$1000$\,V/cm along $\mathbf{H}\parallel x$ and $\mathbf{E}\parallel y$, respectively, which can establish a single AFM domain as demonstrated earlier \cite{Kocsis2018}.

%%%%%%%%%%%
% Results: spectroscopy
%%%%%%%%%%%

We present the absorption spectra of \LCPO/ measured at the base temperature of the optical cryostat, $\sim$5\,K in Fig.~\ref{fig:fig_spectra} (a). The linearly polarized spectra, $\mathbf{E}^{\omega}\parallel x$ ($\mathbf{E}^{\omega}\parallel y$) shown in blue and green (red and orange) correspond to the absorption of the two AFM domains stabilized by ME poling. We measured spectra in all four combinations of the signs of the poling fields, and found that spectra corresponding to the same domain overlap within the noise of the experiment. %In the present experimental setup, the small permanent magnets providing the poling magnetic fields could not be removed after poling, thus, the measurements were carried out in the poling magnetic fields. The electric field was driven to zero before measurements. In accordance with THz spectroscopy \cite{Kocsis2018}, we detected no change in the spectra when the electric field was switched off, nor did we expect any measurable effects of the small magnetic field. 
We observed a series of absorption peaks, which are consistent with the published ones in the region between 1.55\,eV and 3\,eV \cite{Rivera1998}. Two strong and sharp absorption peaks are present at low energies, around $\sim$80\,meV and $\sim$120\,meV, making the sample completely opaque in the vicinity of these resonances.

Fig.~\ref{fig:fig_spectra}. (b) shows the absorption difference $\Delta\alpha$, for the two AFM domains for both light polarizations. This absorption difference is the NDD signal, as measuring the absorption difference for $\pm$$\mathbf{k}$ is equivalent to the measurements on the time-reversed domain pairs with a fixed propagation direction \cite{Kocsis2018,Kocsis2019}. We observed the strongest NDD for $\mathbf{E}^{\omega}\parallel y$ at 0.776\,eV, or 1597\,nm, close to the telecommunication wavelength 1550\,nm. To quantify the difference, we followed Ref.~\cite{Kimura2024}, which results $\Delta\alpha/\alpha_0 = 34\%$ at 1597\,nm, where $\alpha_0=\alpha_{\mathrm{dom1}}+\alpha_{\mathrm{dom2}}$.

In agreement with former publications \cite{Rivera1998,Kornev1999,Kimura2024}, we assigned the observed absorption peaks to crystal field excitation of Co$^{2+}$ ions occupying a distorted octahedral environment. The energy level scheme proposed in Ref.~\cite{Kornev1999} based on crystal-field calculations accurately predicts the positions of the lowest-lying transitions as shown in Fig.~\ref{fig:fig_spectra}. These transitions within the crystal field split $^{4}$F term at 0.08, 0.16, 0.64, 0.68, 0.95, 1.59 eV correspond well to the absorption peaks at 0.08, 0.12, 0.6, 0.7, 0.95 and 1.6\,eV, though we cannot exclude that any of the two lowest-lying excitations (0.08 and 0.12\,eV) might be assigned to an infrared active phonon mode. The weaker features observed around 0.25 and 0.4\,eV deviate from the crystal field calculations, thus, they may originate from impurities. Transitions to the higher-lying $^{4}$P and $^{2}$G terms are %\textcolor{red}{predicted to be at 2.22, 2.23, 2.51, 2.9(4), 2.9(7), 2.34 , 2.47, 2.53(1), 2.53(3), 2.57, 2.58 and 2.64 eV}, which is 
consistent with the series of resonances observed in the region between 2 and 2.6\,eV. However, assigning the specific fine structure within this range is challenging due to overlapping transitions and the multitude of closely spaced excitations predicted by the crystal field model.

The site symmetry group of Co$^{2+}$, $C_\mathrm{S}$, which contains a single mirror plane normal to the $y$ axis, enforces a selection rule as pointed out by Kornev \textit{et al.}~\cite{Kornev1999}. In the weak spin-orbit limit, the wavefunctions can be labelled by a parity quantum number: they either preserve or change signs under mirror symmetry, i.e.,~they transform as the $A^\prime$ or $A^{\prime\prime}$ irreducible representations of $C_\mathrm{S}$, respectively. When the electric field oscillates along the $\mathbf{E}^{\omega}\parallel y$ direction, -- thus, it transforms as $A^{\prime\prime}$, -- it excites transitions between states with different parity, whereas, electric fields oscillating in the perpendicular directions induce parity-conserving transitions as these field components correspond to $A^{\prime}$. According to the calculations, the transitions at 0.08, 0.6 and 1.6\,eV are active for $\mathbf{E}^{\omega}\parallel x$, while the resonances at 0.12, 0.7 and 0.95\,eV are excited by $\mathbf{E}^{\omega}\parallel y$. Besides the two lowest-lying excitations with too high intensity to resolve, the other four NIR excitations mostly follow this electric dipole selection rule, namely, their intensity is stronger for the allowed direction. However, the cancellation is not complete for the orthogonal direction, and the absorption peak at 0.7\,eV has a large spectral weight also for $\mathbf{E}^{\omega}\parallel x$, which may indicate spin-orbit mixing.

\begin{figure}[h]
\centering
\includegraphics[width = \linewidth]{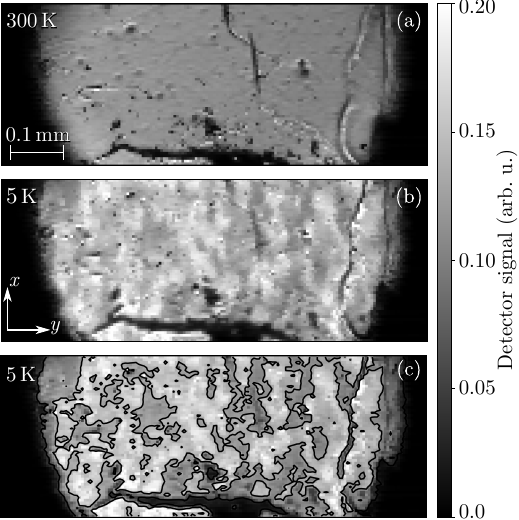}
%\captionsetup{justification=Justified}
\caption{Transmission images of \LCPO/ recorded at (a) 300\,K and (b), (c) below $T_{\mathrm{N}}$ at 5\,K after zero-field cooling (ZFC). (b) Brighter and darker regions indicate different antiferromagnetic domains. (c) The domain walls are highlighted by black lines.}
\label{fig:fig_domains}
\end{figure}

The crystal field excitations of Co$^{2+}$ in this non-centrosymmetric local environment are ideal for hosting non-vanishing NDD emerging at simultaneously electric and magnetic dipole active transitions \cite{Barron2004book,Jung2004,Saito2008}. 
$E^{\omega}_x$ ($E^{\omega}_y$) and $H^{\omega}_y$ ($H^{\omega}_x$) belong to the same irreducible representations $A^{\prime}$ ($A^{\prime\prime}$) of the site symmetry, thus, when a transition is electric dipole allowed the corresponding magnetic field component can also excite it. Furthermore, the electric and magnetic field components transforming the same way are perpendicular to each other just like the oscillating fields of light enabling both of them to couple to the transitions in the studied geometry. These symmetry arguments explain the appearance of NDD in \LCPO/, however, to reproduce the absorption strengths and the magnitude of the NDD microscopic model calculations including spin-orbit coupling are needed, which is beyond the scope of this paper.

%%%%%%%%%%%
% Results: microscopy
%%%%%%%%%%%

Next, we utilized the absorption contrast of the domains, the NDD for imaging. Since $\Delta\alpha$ is the largest around 0.8\,eV, we used a 1550\,nm (0.8\,eV) laser diode as a light source and polarized its light to $\mathbf{E}^{\omega}\parallel y$ to enhance the contrast. The light beam was focused onto the sample and then collected with long working-distance objectives (MY10X-823). The achieved spot size was 4\,$\mu$m. Behind the sample, an InGaAs detector measured the transmitted intensity. The pixels of the image were obtained as the sample moved in the focus in a raster fashion. The slow drift of the detector signal was compensated in each row by subtracting the first datapoint, where the aperture of the sample holder covered the laser beam.

The images collected at 300\,K and at 5\,K after zero-field cooling are shown in Fig.~\ref{fig:fig_domains}. Besides the topographic features observed at room temperature [see Fig.~\ref{fig:fig_domains} (a)] higher and lower intensity regions corresponding to the AFM domains emerge in the magnetically ordered phase as shown in Fig.~\ref{fig:fig_domains} (b). The AFM domains are highlighted in Fig.~\ref{fig:fig_domains} (c) by marking the intensity changes. The typical domain size is a few 10\,$\mu$m, smaller than formerly reported values in Ref.~\cite{VanAken2007,Zimmermann2009,Zimmermann2014}.

The transmission microscopy experiments revealed its capability to map the two AFM domains in the bulk crystal, offering a simpler approach compared to surface sensitive SHG imaging. We note, however, that SHG has the advantage that it probes a three-index tensor, thus, it provides more information on the magnetic symmetry. Indeed, some elements of the SHG tensor forbidden in $mmm$' become finite, though weak below $T_\mathrm{N}$\cite{VanAken2007,Zimmermann2009,Zimmermann2014}. These finite elements indicate that the actual symmetry might be lower, in agreement with neutron scattering \cite{Vaknin2002}. The symmetry reduction would also lead to an increase in the number of the domains, however, we could not unambiguously distinguish them by NDD microscopy.

\begin{figure}[h]
\centering
\includegraphics[width = \linewidth]{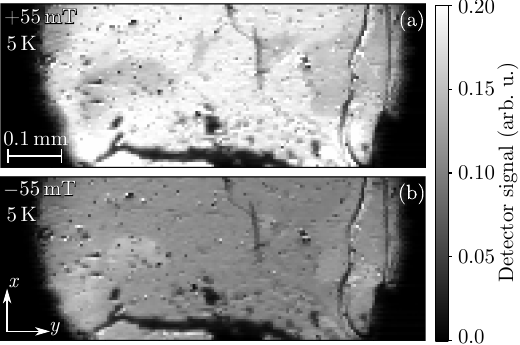}
%\captionsetup{justification=Justified}
\caption{ Transmission images of \LCPO/ recorded at 5\,K after cooling the sample only in a magnetic field. (a) and (b) corresponds to images measured after poling only with magnetic fields pointing along and opposite to $\mathbf{H}\parallel y$, respectively.}
\label{fig:fig_domains_B}
\end{figure}

We also recorded intensity images after poling in magnetic fields only. The field was applied along $\mathbf{H}\parallel y$ and its magnitude was 55\,mT. The images measured in zero-field after poling in +55\,mT and -55\,mT are shown in Fig.~\ref{fig:fig_domains_B}(a) and \ref{fig:fig_domains_B}(b), respectively. Surprisingly, a large area of the sample was turned into a single AFM domain corresponding to bright and dark area in Fig.~\ref{fig:fig_domains_B}(a) and \ref{fig:fig_domains_B}(b), respectively. In smaller patches, the intensity change is smaller suggesting that the sample is not in a monodomain state over the entire thickness. This partial control of AFM domains by magnetic poling is consistent with the findings of Rivera \cite{Rivera1994}, where butterfly-shaped $\mathbf{P}$-$\mathbf{H}$ loops were observed close to $T_\mathrm{N}$. In our case, surface charge accumulation can be discarded as the sample is directly grounded by the DC power source through electrical contacts. Neutron diffraction experiments suggested that weak ferromagnetism can be consistent with the small rotation of the moments away from the $y$ axis and it might be responsible for magnetic field effects \cite{Vaknin2002}. However, if spatial inversion is a symmetry of the paramagnetic state, linear coupling between the axial vector of the ferromagnetic moment and the inversion-symmetry-breaking AFM order parameter is not possible, leaving this issue open for further investigations.

In summary, we studied the absorption spectrum of \LCPO/ in its antiferromagnetically ordered phase over the infrared and visible spectral ranges. We observed a significant absorption difference between the two antiphase AFM domains reaching 34\,\%
at 0.776\,eV (1597\,nm). Notably, this large zero-field NDD arises from crystal field excitations of Co$^{2+}$ ions near the telecommunication wavelength of 1550\,nm. Our experiments broaden the number of transition metal ions enabling large NDD, which may become applicable in non-reciprocal optical components like isolators or switchable optical diodes at these technologically relevant wavelengths \cite{Kezsmarki2015, Kimura2024}.  We utilized this spontaneous optical anisotropy for visualizing AFM domains. Since this is based on simple transmission microscopy, this approach offers a unique possibility to map AFM domains in various ME materials. 
 
\begin{acknowledgments}
The Authors are grateful to Heung-Sik Kim and Istv\'an K\'ezsm\'arki for valuable discussions. This work was supported by the Hungarian National Research, Development and Innovation Office–NKFIH Grants No. FK 135003, by the Ministry of Culture and Innovation and the National Research, Development and Innovation Office within the Quantum Information National Laboratory of Hungary (Grant No. 2022-2.1.1-NL-2022-00004). V. K. was supported by the Alexander von Humboldt Foundation.
\end{acknowledgments}

\bibliography{LCPO_bib}

\end{document}